\documentclass[prl,aps,twocolumn,showpacs,floatfix,amsmath,superscriptaddress,amssymb]{revtex4}
\usepackage{latexsym}
\usepackage{graphicx}
\usepackage{dcolumn}
\usepackage{bm}
\usepackage{psfrag}
\usepackage{subfigure}

\newcommand{\be}{\begin{equation}}
\newcommand{\ee}{\end{equation}}
\newcommand{\ba}{\begin{eqnarray}}
\newcommand{\ea}{\end{eqnarray}}
\newcommand{\baa}{\begin{eqnarray*}}
\newcommand{\eaa}{\end{eqnarray*}}

\def\be{\begin{equation}}
\def\ee{\end{equation}}
\def\ba{\begin{eqnarray}}
\def\ea{\end{eqnarray}}

\def\LSCO{La$_{2-x}$Sr$_x$CuO$_4$}

\def\C60{A$_x$C$_{60}$}

\def\HgCu3{HgCa$_2$Cu$_3$O$_{8+y}$}
\def\HgCu4{HgBa$_2$Ca$_3$Cu$_4$O$_{10+y}$}
\def\TlCu{Tl$_2$Ba$_2$CuO$_{6+\delta}$}
\def\TlCu3{Tl$_2$Ba$_2$Ca$_2$Cu$_3$O$_{10+y}$}
\def\TlCu4{Tl$_2$Ba$_2$Ca$_3$Cu$_4$O$_{12+y}$}

\def\BiCu3{Bi$_2$Sr$_2$Ca$_{2}$Cu$_3$O$_y$}

\def\8LSCO{La$_{1.88}$Sr$_{.12}$CuO$_4$}
\def\110LNSCO{La$_{1.5}$Nd$_{0.4}$Sr$_{0.1}$CuO$_{4}$}
\def\stage4LCO{La$_{2}$CuO$_{4+\delta}$}
\def\Y248{YBa$_2$Cu$_4$O$_8$}

\def\NbSe2{NbSe$_2$}
\def\TaSe2{TaSe$_2$}
\def\TiSe2{TiSe$_2$}

\begin{document}

\title{Hysteresis and Noise from Electronic Nematicity \\
in High Temperature Superconductors}

\author{E. W. Carlson}
\affiliation{Department of Physics, Purdue University, West Lafayette, Indiana 47907, USA}
\author{K. A. Dahmen}
\author{E. Fradkin}
\affiliation{Department of Physics, University of Illinois, Urbana, Illinois  61801, USA} 
\author{S. A. Kivelson}
\affiliation{Department of Physics, Stanford University, Stanford, California 93105, USA}
\affiliation{Department of Physics and Astronomy,
University of California, Los Angeles, California 90095, USA}

\date{\today}

\begin{abstract}
An electron nematic is a translationally invariant state which spontaneously breaks the discrete rotational 
symmetry of a host crystal.  In a clean square lattice, the electron nematic has two preferred orientations, 
while dopant disorder favors one or the other orientations locally. In this way, the electron nematic in a 
host crystal maps to the random field Ising model (RFIM). 
Since the electron nematic has anisotropic conductivity, 
we associate each Ising configuration with a resistor network, and use what is known about the RFIM to 
predict new ways to test for electron nematicity using noise and hysteresis.  In particular, we have 
uncovered a remarkably robust linear relation  between the orientational order and the resistance anisotropy 
which holds over a wide range of circumstances.

\end{abstract}

\pacs{72.70.+m, 74.25.Fy, 75.10.Nr}

\maketitle 

In the high temperature superconductors, in addition to superconductivity, 
there may exist various other types
of order which break spatial symmetries of the underlying crystal, 
especially in the ``pseudogap'' regime
at low doping.\cite{kivelson-fradkin-emery-nature-98,sudip,rmp,concepts,varmaloops}  
However, it is often surprisingly difficult to 
obtain direct experimental evidence which permits one to clearly 
delineate in which 
materials, and in what range of temperature and doping, these phases occur.  
One such candidate order is the electronic nematic, which breaks orientational, 
but not translational, symmetry.\cite{kivelson-fradkin-emery-nature-98}
Orientational long range order (LRO) induces
transport anisotropy, since it is easier to conduct along one direction of the
electronic nematic than the other, and this is a natural 
way to look for nematic order.
On the other hand, quenched disorder couples linearly to the order parameter, 
and so, for quasi-2D systems, will typically change a true thermodynamic transition into a mere crossover.  

In this letter, we propose methods for detecting local electronic nematic order (nematicity)
even in the presence of quenched disorder.
We show that an electron nematic in a disordered square lattice 
maps to
a random field Ising model (RFIM)\cite{nattermann-97} and that the
transport properties of the system can be determined from a random resistor network
whose local resistances reflect the local nematic transport anisotropy.
In a host crystal such as the cuprates, an electron nematic
(which may arise from local ``stripe'' correlations) tends to lock 
to favorable lattice directions, often 
either ``vertically'' or ``horizontally'' along Cu-O bond directions.
The two possible orientations can be represented
by an Ising pseudospin, $\sigma=\pm 1$,\cite{abanov} 
and the tendency of neighboring nematic
patches to align corresponds to a ferromagnetic interaction.
\begin{figure}[tb]
\begin{center}
\subfigure[Nematic patches]{
\includegraphics[width=.45\columnwidth]{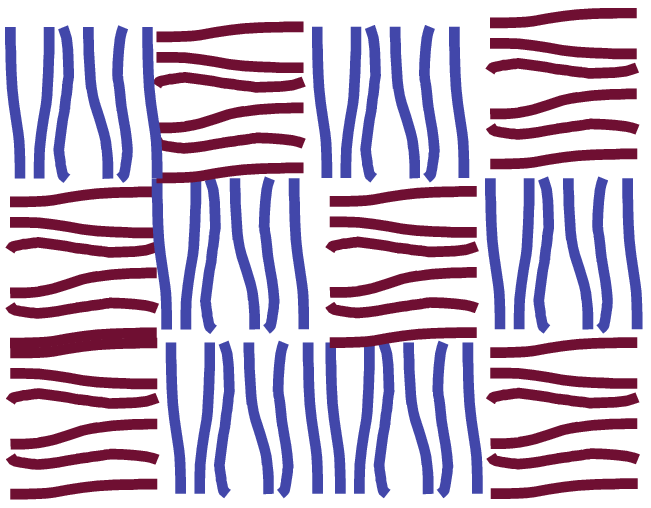}
\label{fig:stripe-domains}}
\subfigure[Resistor Network]{
\includegraphics[width=.45\columnwidth]{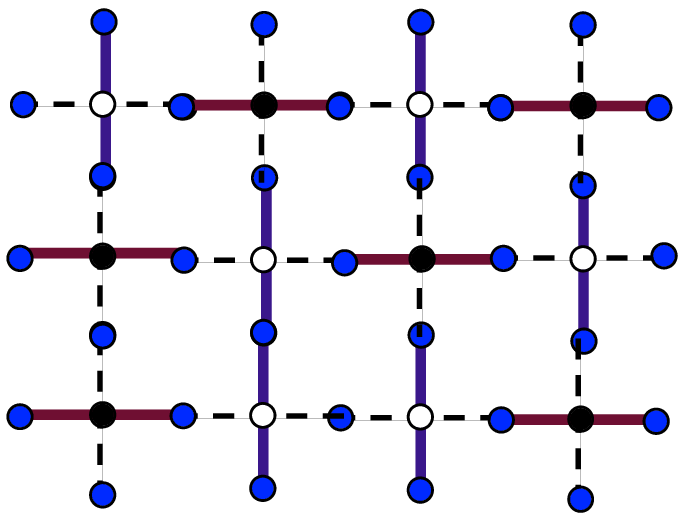}
\label{fig:resistor-network}}
\end{center}
\caption{Mapping of nematic patches to random resistor networks.  
(a) Configurations of nematic  patches generated by the random field 
Ising model.  
(b) Corresponding resistor network, modeling local anisotropic conduction 
in each nematic patch. 
Solid lines are small resistors, and dotted lines are large ones.}
\label{fig:mapping}
\end{figure}
In any given region, disorder due to dopant atoms between the cuprate planes
produces electric field gradients which locally favor one orientation
or the other, and act like a random field on the electronic nematic,
as illustrated in Fig.~\ref{fig:stripe-domains}.
Thus an electron nematic in a host crystal maps to the RFIM:
\begin{equation}
H = -J \sum_{<i,j>}\sigma_i \sigma_j  -\sum_i (h+h_i) \sigma_i 
\end{equation}
where $J>0$ is the coupling between neighboring nematic patches.
The local disorder field $h_i$
is taken to be gaussian, with a disorder strength $\Delta$ characterized 
by the width of the gaussian
distribution.
The symmetry breaking field $h$ 
may be produced by, {\em e.g.}, uniaxial strain, 
high current,\cite{reichhardt-05} or magnetic field. 
For example, nematics can be aligned by an external magnetic field due to diamagnetic 
anisotropy, in which case $h \propto o(H^2)$.\cite{chaikin-lubensky}

The macroscopic resistance anisotropy of a
nematic\cite{fradkin-kivelson-99,Cooper-01,ando} 
transforms under rotations in the same way as the orientational
order parameter $m=(1/N)\sum_i \sigma_i$ and
is a natural candidate for measuring nematic order.
The (normalized) resistance anisotropy is
$R_a \equiv \left(\frac{r+1}{ r-1}\right)\left(\frac{R_{xx} - R_{yy}}{R_{xx}+R_{yy}}\right)$,
where $r \equiv R^{\rm max}_{xx}/R^{\rm min}_{xx}$
is the ratio of the extremal macroscopic resistances in the two fully oriented states.
To obtain the transport properties, we map each pattern of local nematic 
orientations generated by Monte Carlo simulations of the RFIM 
(see Fig.~\ref{fig:stripe-domains}) to a resistor network
(see Fig.~\ref{fig:resistor-network}) which models the
local anisotropic transport in each nematic patch.
Each nematic patch (Ising pseudospin) becomes one node in the 
resistor network, with four
surrounding resistors determined by the nematic orientation.
For a ``vertical'' nematic patch, we assign the resistors to the ``north'' 
and ``south''
to be small, $R_{\rm small}$, while the resistors to the ``east'' 
and ``west'' 
are large, $R_{\rm large}= r R_{\rm small}$.  For a ``horizontal'' nematic patch, 
these assignments are reversed.  
When all patches are fully oriented
so that $m\rightarrow 1$, the macroscopic 
resistance anisotropy must also saturate, $R_a \rightarrow 1$.  
More generally, in the thermodynamic limit
$<R_a> = m F(m,T,r)$, where $F$ is an even function of $m$.
Remarkably, as we will see below, $F=1$ to a 
very good approximation and, under
a wide range of circumstances, $<R_a> \approx m$ throughout the entire range of $m$.  


\begin{figure}[tb]
\begin{center}
\subfigure[$R_a$ {\em vs.} $h$.]{
\psfrag{R}{$\;\;\;\;R_a$}
\psfrag{H}{$\;\;\;\;\;\;\;h$}
\psfrag{0}{$\!\!0$}
\psfrag{-0.5}{}
\psfrag{0.5}{}
\psfrag{-1}{$\!\!-1$}
\psfrag{1}{$\!\!1$}
\psfrag{-4}{$-4$}
\psfrag{4}{$4$}
\includegraphics[width=.47\columnwidth]
{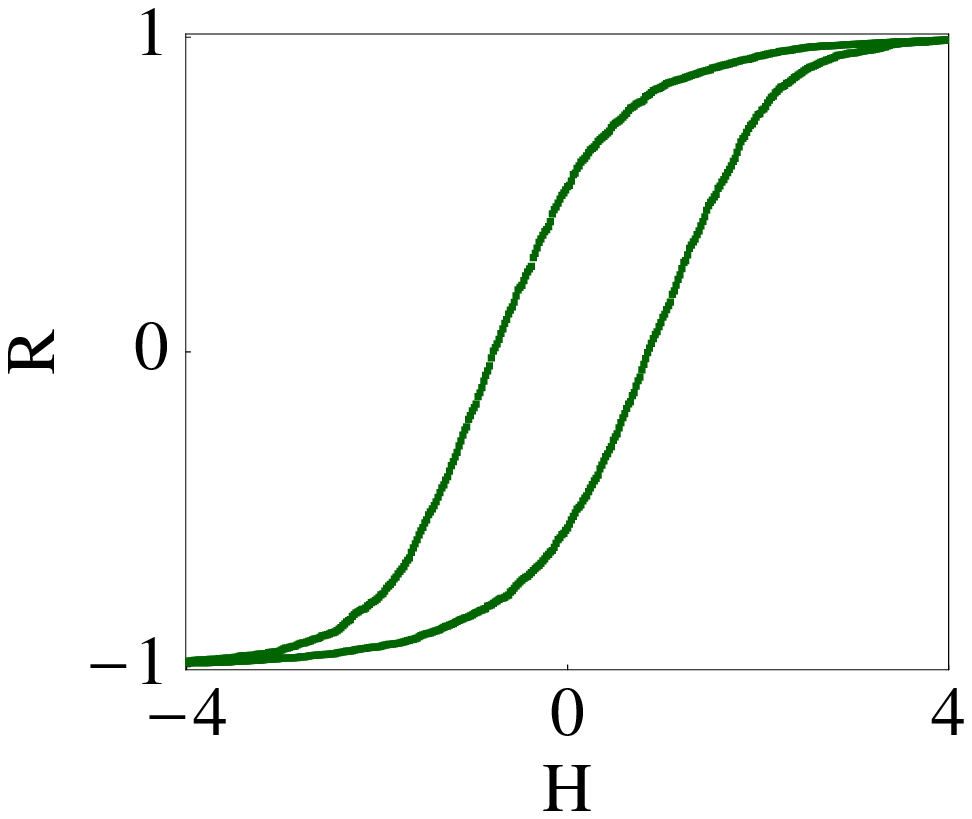}\label{fig:hysteresis}}
\subfigure[$R_a$ vs. $m$]{
\psfrag{M}{$\;m$}
\psfrag{R}{$R_a$}
\psfrag{0}{$0$}
\psfrag{-0.5}{}
\psfrag{0.5}{}
\psfrag{-1}{$\!\!-1$}
\psfrag{1}{$\!\!1$}
\includegraphics[width=.47\columnwidth]
{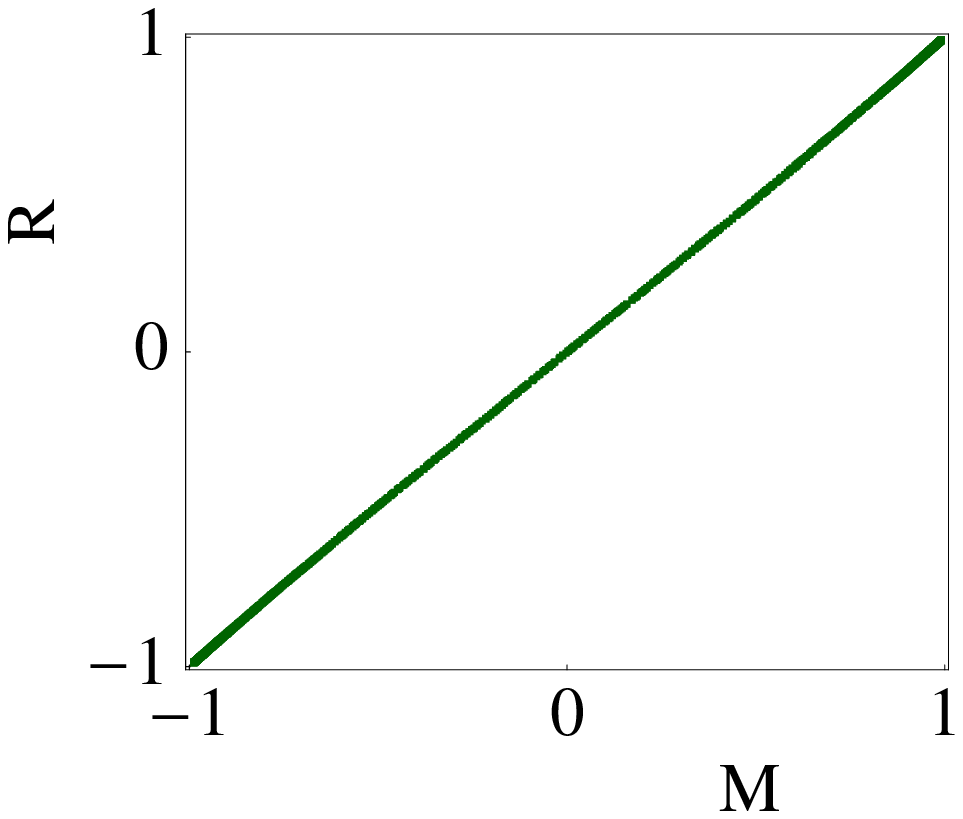}\label{fig:linear}}
\end{center}
\caption{Hysteresis and comparison of 
the macroscopic resistance anisotropy
to the orientational order for 
size $L\times L=100\times 100$, disorder strength $\Delta=3 J$, temperature $T=0$,
microscopic resistance anisotropy $r = 2$,
and external field sweep rate $\Omega = J/N$, where $N=L \times L$.
{(a) Hysteresis}
of the resistance anisotropy
$R_a$ (see text)
{\em vs.} the symmetry breaking field $h$.  (b) Resistance anisotropy $R_a$ {\em vs.} 
orientational order parameter $m$.  
The resistance anisotropy is a remarkably good
indicator of orientational order.
}
\label{fig:Hysteresis-100}
\end{figure}
We use a Glauber update method to generate configurations of the RFIM,
with periodic boundary conditions on the pseudospin lattice.  The details of
the algorithm are contained in Ref.\cite{white-travesset-05}.
We then calculate the resistance anisotropy $R_a$ of the corresponding 
resistor network by the  following method:
For $R_{xx}$, we assign a uniform applied voltage to every site at the 
far left end of the lattice, and a uniform ground 
to the far right end of the lattice, with open voltage 
boundary conditions
in the $y$ direction, corresponding to
infinite resistors coming out of the top and bottom edges
of the network.  We then apply the bond propagation algorithm\cite{lobb} 
to reduce the network
to a single resistor equivalent to the macroscopic $R_{xx}$.
$R_{yy}$ is similarly calculated from the same starting resistor network, 
but with
boundary conditions appropriate for $R_{yy}$.

Below a critical disorder strength, the 3D RFIM possesses
a finite temperature phase transition to a 
low temperature ordered phase.  However, in 2D the critical disorder is zero, and LRO
is forbidden.\cite{imry-ma} 
The quasi-2D case of coupled RFIM 
planes also has finite critical disorder,\cite{zachar-03}
but it is exponentially small.
We focus here on two dimensions, and demonstrate that even in this 
case, where orientational LRO 
is forbidden, it is possible to detect the proximity to order through noise 
and hysteresis measurements.


We first present results at zero temperature, which exhibits the
nonequilibrium behavior associated with hysteresis.
Fig.~\ref{fig:Hysteresis-100} shows a simulation of the RFIM for a 
system of size 
$L=100\times 100$ at zero temperature.  We present results for 
which the field $h$ 
is incremented at a sweep rate
$\Omega = J/N$, where $N=L\times L$ is the system size and $h = \Omega t$.  
In Fig.~\ref{fig:hysteresis}, we show a hysteresis loop 
for the resistance anisotropy $R_a$ 
{\em vs.} the symmetry breaking field $h$, starting from the
fully oriented state at $h=-\infty$.  
Fig.~\ref{fig:linear} plots $R_a$ {\em vs.} $m$ through one cycle of
the hysteresis loop. 
The relationship is remarkably linear, 
and the macroscopic resistance anisotropy $R_a$ shows precisely the same hysteretic  
behavior as $m$.
This linear relation throughout the cycle is surprising, 
since for a given magnetization $m$ not too close to $1$,
there are several values possible for the macroscopic resistance anisotropy.
This is because the flow of current through the random resistor network depends 
not only on the relative concentration of ``up'' and ``down'' resistor nodes,
but also on their spatial relation.
In fact, for $m=0$, $R_a$ can take values between 
$\pm (r^2-1)/(r^2+6r+1)$.
However, for typical configurations generated by the RFIM,
the main contribution to $R_a$ is controlled by 
the number of up and down nodes,
with their spatial relation being a small effect which goes to zero 
as the system size $L\rightarrow \infty$ or as $r \rightarrow 1$.
(We do observe variations in $R_a$ between configurations with the same magnitude
of $m$ at finite temperature or for small system sizes.)
For the same reason, 
in confined geometries there can be more noise in the resistance anisotropy than is present 
in the true orientational order parameter.  
The fact that $<R_a> \approx m$ over a wide range of parameters
means that noise and hysteresis in this measurable property can be
used to detect the electronic nematic.


\begin{figure}[tb]
\psfrag{m}{$m$}
\psfrag{h}{$h$}
\begin{center}
\subfigure[Hysteresis]{
\psfrag{A}{\small{$A$}}
\psfrag{B}{\small{$B$}}
\includegraphics[width=.48\columnwidth]
{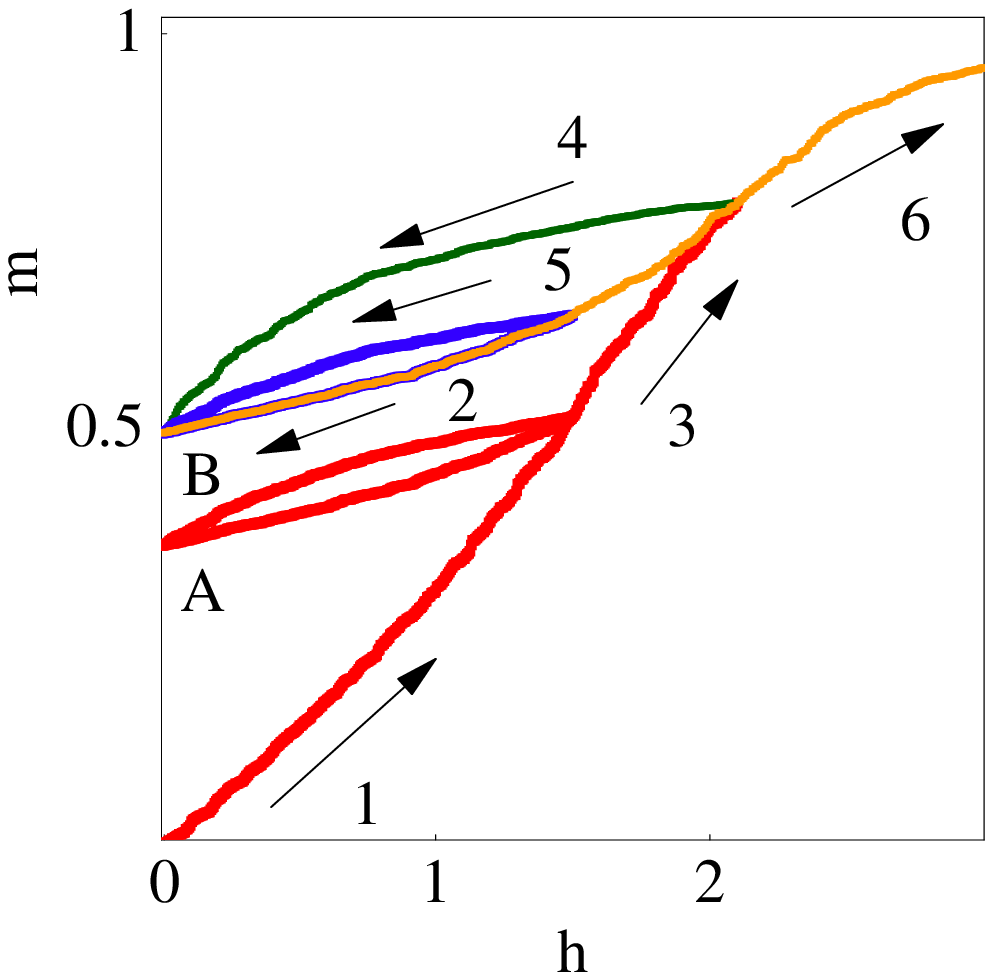}}
\subfigure[Subloops]{
\psfrag{A}{$A$}
\psfrag{B}{$\!\!B$}
\includegraphics[width=.48\columnwidth]
{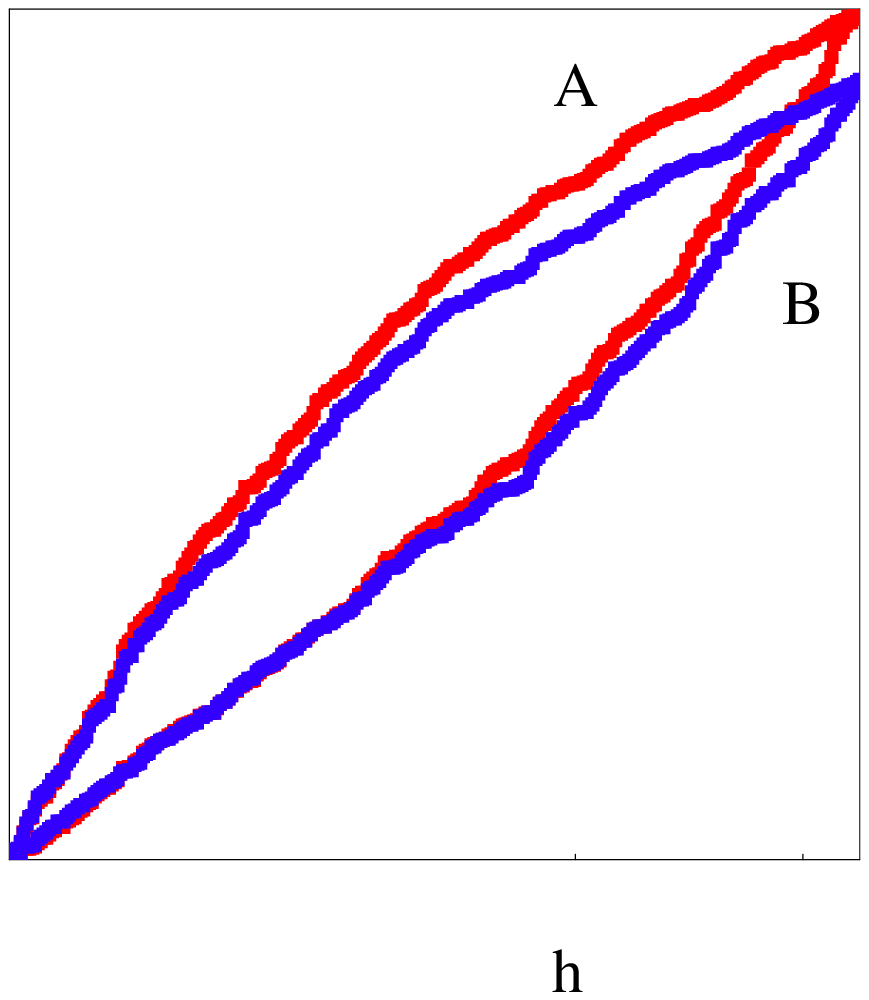}\label{fig:inset}}
\end{center}
\caption{
(a) Field ramp-up from a disordered configuration at $T=0$, with hysteresis
subloops, for system size $L\times L=100 \times 100$ 
and disorder strength $\Delta = 3 J$. 
The arrows indicate the order in which the external field sweep is taken.
Subloops $A$ and $B$ 
are both taken between the same extremal orienting fields $h=1.5 J$ and $h=2.1 J$, 
but with different histories.  
(b) Incongruency of subloops $A$ and $B$, due to the presence of interactions.   
Here the orientational order parameter
has been shifted to compare the shape of the two subloops.
Subloop $A$, executed at lower $m$, has a higher slope than 
subloop $B$, executed at higher $m$.
}
\label{fig:subloops}
\end{figure}
Hysteresis subloops can be used to 
determine qualitatively the relative importance of interactions and disorder.  
Fig.~\ref{fig:subloops} shows the behavior of subloops in the hysteresis curve
of orientational order $m$ {\em vs.} orienting field $h$.
Starting from zero field and a thermally disordered configuration, 
the field is swept up along path $1$, then switched back along path $2$
to take subloop $A$ before continuing along path $3$.  Then path $4$
begins a second subloop, within which path $5$ begins subloop $B$, before
continuing to raise the field through path $6$ until $m$ has saturated,
$m \rightarrow 1$.

Notice that the subloops close, and also that once a subloop has closed, continuing to 
raise the field does not disturb the structure of the outer loop.
This is indicative of return-point memory, a characteristic of the RFIM.\cite{sethna-93}  
Additionally, subloops $A$ and $B$ in Fig.~\ref{fig:subloops}
are a comparison
between the same two field strengths.  The fact that the two are incongruent
as shown in Fig.~\ref{fig:inset}
indicates that interactions are present, 
and possibly even avalanches, 
{\em i.e.} that the hysteresis is not simply a linear superposition of 
elementary hysteresis loops of independent grains as in the Preisach model.\cite{dahmen-review,hallockprl,Mayergoyz-book}  
Here, we have used  a disorder strength of 
$\Delta=3 J > \Delta^{3D}_{crit}= 2.16 J$.
At lower disorder, the subloops become narrower and difficult
to resolve.  Also, as temperature is increased, thermal disorder
means that subloops become narrower, and no longer close precisely.

Recent magnetization measurements 
by Panagopoulos {\it et al.\/}
\cite{panagopoulos-magnetic-04,panagopoulos-04} 
on {\LSCO} reveal hysteretic behavior reminiscent of the RFIM.
Although some of the results look very similar to our Fig.~\ref{fig:subloops}, 
in fact our model does not explicitly include the ferromagnetic moments measured in 
this experiment.  However, the small magnitude of the measured magnetization
may be consistent with 
ferromagnetic moments 
arising at defects in the local striped antiferromagnetic order of a nematic patch.


We now discuss equilibrium fluctuations.  
In confined geometries such as nanowires and dots, this model can exhibit
telegraph noise in $R_a$ due to thermal fluctuations of large correlated clusters.  
Recent transport experiments on underdoped YBCO nanowires by Bonetti {\it et al.\/}\cite{bonetti-04} reveal 
telegraph-like noise in the pseudogap regime. Using a 
YBCO nanowire of size 250nm$\times$500nm, they
found that 
a time trace of the resistance at constant temperature T=$100$K shows
telegraph-like fluctuations of magnitude $0.25\% $,
on timescales on the order of $50$ seconds.  These large scale, slow switches
can be understood within our model as thermal fluctuations of
a single correlated cluster of nematic patches.
Individual nematic patches thermally fluctuating
between two orientations also produce noise, but at higher frequency, and
with smaller effect on the macroscopic resistance. 
By comparing the magnitude of the smallest typical resistance change to the resistance
change at a telegraph noise event,
one can estimate that the correlated cluster in the nanowire
contains at least 5-6 nematic patches.

\begin{figure}[tb]
\psfrag{time}{\small{$t \times 10^3$}}
\psfrag{N}{\small{$N \times 10^3$}}
\psfrag{11}{\small{$\!\! 11$}}
\psfrag{11.5}{}
\psfrag{12}{\small{$\!\! 12$}}
\psfrag{12.5}{}
\psfrag{1000}{}
\psfrag{2000}{}
\psfrag{3000}{}
\psfrag{4000}{\small{$4$}}
\psfrag{4}{\small{$\!\!4$}}
\psfrag{6000}{}
\psfrag{8000}{\small{$8$}}
\begin{center}
\subfigure[$R_{xx}$ {\em vs.} time]{
\psfrag{0}{\small{$0$}}
\psfrag{Rxx}{\small{$R_{xx}$}}
\includegraphics[width=.49\columnwidth]
{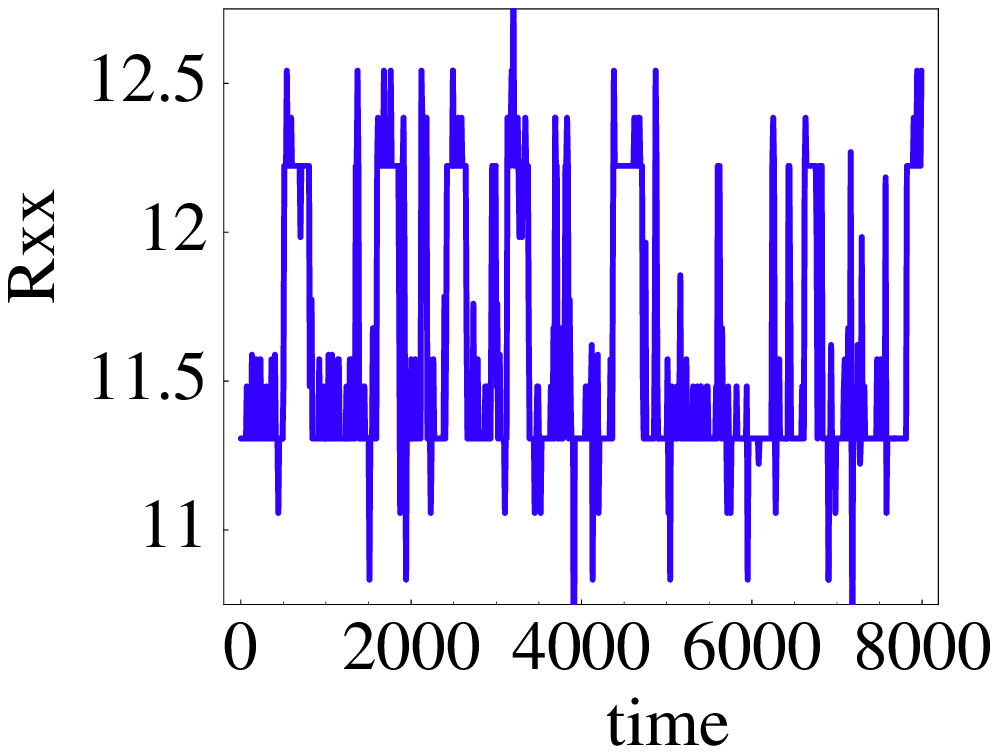}\label{fig:timetrace}}
\subfigure[Histogram]{
\psfrag{0}{\small{$\!\!0$}}
\psfrag{Rxx}{\small{$R_{xx}$}}
\includegraphics[width=.47\columnwidth]
{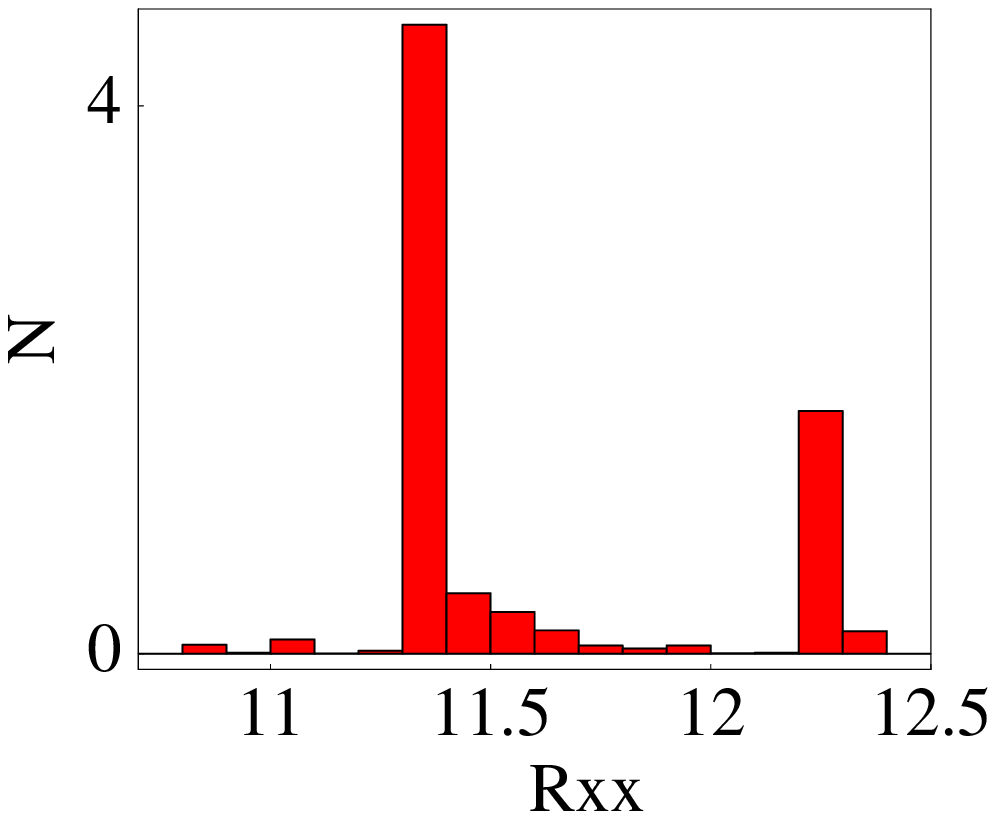}\label{fig:histogram}}
\end{center}
\caption{(a) Telegraph noise in a confined geometry.  
Resistance $R_{xx}$ {\em vs.} time
of a small system of size $L \times L = 6 \times 6$ 
with disorder strength $\Delta=2J$,
at temperature $T=0.5J$, in zero applied field, $H=0$.  
An initial random state is allowed to 
thermalize for 10,000 MC steps before measurements are taken.
The local microscopic conductivity anisotropy in the resistor network is 
$r = 2$. Small fluctuations are due to a single 
nematic patch fluctuating.
Large switches are due to the thermal fluctuations of a single correlated cluster of
nematic patches.
(b) Histogram of the time series.  $N$ is the number of occurrences of a 
particular resistance.    There are two main states, one for which
the correlated cluster is ``up'', and the other for which it is ``down''. }
\label{fig:RvTime}
\end{figure}

The nanowire represents a rather small system size when mapped to the RFIM.
Neutron scattering experiments
on the incommensurate peaks of underdoped YBCO indicate a coherence length 
of roughly $40$nm.  If we take this as an estimate of the size of one nematic
patch
(mapped to a single pseudospin in the RFIM), the nanowire is about 
$6 \times 12$ patches wide, and the effect of flipping a 
correlated cluster of pseudospins can dominate the response.
In Fig.~\ref{fig:timetrace}, we show a time series of $R_{xx}$
in thermal equilibrium ($h=0$)
for a small system of size $L\times L=6 \times 6~$ at finite temperature with
disorder strength $\Delta = 2J$, which is less than the 3D critical value.
Notice the sizable thermal fluctuations in the macroscopic resistance.  
The high frequency noise is due to the thermal 
fluctuations of a single Ising pseudospin. The lower frequency telegraph noise,
in which the resistance changes dramatically, is due to the correlated 
fluctuations of a cluster of pseudospins, in our case a cluster of 
$4$ pseudospins. 
The histogram in Fig.~\ref{fig:histogram}
shows that the system switches between two main states.


In general, correlated clusters will thermally switch with a timescale 
which rapidly increases with their size.  
Confined geometries such as the nanowire are
more likely to have a system-spanning
cluster which causes slow, large scale noise.  For larger system sizes,
the histogram becomes smoother and can have multiple peaks. 
For systems larger than the Imry-Ma correlation length, large scale switches 
are highly unlikely.  A probability distribution of size of cluster {\em vs.} 
frequency can be obtained from a histogram of the 
{\em differential resistance},
in which large scale switches will always be relegated to the tails of the 
distribution.

A further way to test for nematic patches is to measure cross correlations 
between the thermal noise in $R_{xx}$ and $R_{yy}$ as a function of time.\cite{weissman-bismuth}  
This can distinguish between macroscopic resistance fluctuations due to, 
{\em e.g.},
fluctuation superconductivity (which would cause {\em correlated} 
fluctuations of $R_{xx}$ and $R_{yy}$), or due to RFIM-nematic physics
(which would cause {\em anticorrelated} fluctuations). 

This general method of connecting local anisotropic properties of an 
electronic nematic
to macroscopic behavior can be extended to many experimental measurements.
For example, applying an in-plane magnetic field 
should drive 4-fold symmetric incommensurate spin peaks in neutron scattering
into a 2-fold symmetric pattern, in a hysteretic manner as the field is 
rotated from one Cu-O
direction to the other. Superfluid density anisotropy
should also display similar hysteresis with in-plane field orientation.
Furthermore, the presence of correlated clusters in the RFIM 
has implications for STM.
Whereas small correlated clusters have faster switching  dynamics, 
large clusters are much slower,\cite{dahmen-review} so that different size clusters 
have different local power spectra. 
STM can be used to do {\em scanning noise spectroscopy},\cite{stm-noise} 
by measuring the power spectrum as a function of position, in order to
produce a spatial map of correlated nematic clusters.  

In conclusion, we have mapped the electron nematic in a host crystal to the 
random field Ising model.
Using a further mapping to a random resistor network, we have predicted new 
ways to 
detect the electron nematic in disordered systems.  
We have demonstrated 
that the
macroscopic resistance anisotropy is a good measure of orientational order
and 
is expected to display hysteresis and thermal noise characteristic of the 
random field Ising model.
Recent experiments on noise in YBCO nanowires 
and hysteresis in 
LSCO exhibit behavior reminiscent of this model.


We thank T.~Bonetti, D.~Caplan,  D.~Van~Harlingen, C.~Panagopoulos 
and M.~Weissman for many illuminating conversations which motivated this work 
and for sharing their unpublished work,
and we thank T.~Datta and J.~C.~Davis for helpful discussions. 
This work was supported in part through The Purdue Research Foundation 
(EWC),
and by NSF grants
DMR 03-25939 and DMR 03-14279 at UIUC, 
PHY 99-07949 at the 
KITP-UCSB (KAD), 
DMR 04-42537 at UIUC (EF), and DMR-04-21960 at
Stanford and UCLA (SAK). 
 
\bibliography{rfim.bib}
\bibliographystyle{forprl.bst}

\end{document}